\documentstyle[twocolumn,prl,aps,epsfig,amssymb]{revtex}

\def\beqn{\begin{eqnarray}}
\def\eeqn{\end{eqnarray}}
\relax
\newcommand{\ba}[1]{\begin{array}{#1}}
\newcommand{\zpr}{\mbox{$Z^{\prime}$}}
\newcommand{\upr}{\mbox{$U(1)^{\prime}$}}

\newcommand{\mz}{\mbox{$M_Z$}}

\newcommand{\x}{\mbox{$\times$}}
\def\ea{\end{array}}

\def\beq{\begin{equation}}
\def\eeq{\end{equation}}
\def\bea{\begin{array}}
\def\eea{\end{array}}

\def\to{\rightarrow}

\def\[{\left[}
\def\]{\right]}
\def\({\left(}
\def\){\right)}

\def\gm2{{g_\mu -2}}

\def\sm0{{\widetilde{m}_0}}

\def\U1em{{U(1)_{\rm em}}}
\def\to{\rightarrow}

\def\sq2{\sqrt{2}}

\def\End{\end{document}}

\begin{document}

\draft

\twocolumn[\hsize\textwidth\columnwidth\hsize\csname
@twocolumnfalse\endcsname

\setcounter{footnote}{0}

\title{ Electroweak Baryogenesis in a Supersymmetric $U(1)^{\prime}$
Model  }
\author{
 Junhai Kang\,$^1$,~
Paul Langacker\,$^1$,~
 Tianjun Li\,$^2$ ~and~
 Tao Liu\,$^1$
}
\address{
\vspace*{2mm} $^1$ Department of Physics and Astronomy,
University of Pennsylvania, Philadelphia, PA 19104-6396, USA \\
$^2$ School of Natural Science, Institute for Advanced Study,
             Einstein Drive, Princeton, NJ 08540, USA\\
} \maketitle
\begin{abstract}
\hspace*{-0.2cm}
 We construct an anomaly free supersymmetric \upr \ model
with a secluded $U(1)^{\prime}$-breaking sector. We study the
one-loop effective potential at finite temperature,
 and show that there exists a strong enough
first order electroweak phase transition for electroweak
baryogenesis (EWBG) because of the large trilinear term $A_h h S
H_d H_u$ in the tree-level Higgs potential. Unlike in the
MSSM, the lightest stop can be very heavy. We consider the
non-local EWBG mechanism in the thin wall regime, and find that
within uncertainties the observed baryon number can be generated
from the $\tau$ lepton contribution,
with the secluded sector playing an essential role. 
The chargino and neutralino contributions 
and the implications
for the \zpr \ mass and electric dipole moments are briefly
discussed.
\\
{PACS numbers:\,12.60.Jv,\,12.60.Cn\, \hfill [ UPR-1063-T ] }

\end{abstract}
\vskip 1pc]

\setcounter{footnote}{0}
\renewcommand{\thefootnote}{\arabic{footnote}}

The baryon asymmetry of the universe has been measured by
WMAP~\cite{WMAP}. Combining their data with other CMB and large
scale structure results, they obtain
the ratio of baryon density $n_B$ to entropy density $s$
\begin{eqnarray}
n_B/s \sim 8.7 ^{+0.4} _{-0.3} \times 10 ^{-11}~.~\,
\end{eqnarray}

To generate the baryon asymmetry, the
Sakharov criteria~\cite{Sak} must be satisfied:
(1) Baryon number ($B$) violation; (2) $C$ and $CP$ violation;
(3) A departure from thermal equilibrium.
Electroweak (EW) baryogenesis is especially interesting
 because the
 Sakharov criteria can be satisfied in the
Standard Model (SM)~\cite{RTWB}.
However, in the SM the electroweak phase transition (EWPT) cannot be
strongly first order for the experimentally allowed Higgs mass, and the
 CP violation from the CKM matrix is too small.
In the Minimal Supersymmetric
Standard Model (MSSM),
although there are additional sources of CP violation in
 the supersymmetry breaking parameters,
a strong enough first order EWPT requires that the lightest stop
quark mass be smaller than the top quark mass $\sim 175$ GeV.
Also, the mass of the lightest CP even Higgs must be smaller than
120 GeV, which leaves a small window above the current
limit~\cite{MSSMEWBG}. In the Next to Minimal Supersymmetric
Standard Model (NMSSM), a trilinear term $A_h h S H_d H_u$ in the
tree-level Higgs potential may induce a strong enough first order
EWPT~\cite{Pietroni,DFMHS}, and the effective $\mu$ parameter is
given by $ h \langle S \rangle$ from the Yukawa term $h S H_d H_u$
in the superpotential in the best-motivated versions. However,
most versions either involve a discrete symmetry and serious
cosmological domain wall problems~\cite{domain},
or reintroduce the $\mu$ problem~\cite{DFMHS}.

The possibility of an extra \upr \ gauge symmetry is
well-motivated in superstring constructions~\cite{string}.
Similar to the NMSSM, an extra
\upr \  can provide an elegant solution to the $\mu$
problem due to the Yukawa term $h S H_d H_u$~\cite{muprob1,muprob2}.
However,  there are no discrete
symmetries or domain wall problems.
The MSSM upper bound of \mz \ on the tree-level mass of
the lightest MSSM Higgs scalar is relaxed, both in
models with a \upr \ and in the NMSSM, because of the Yukawa term
$h S H_d H_u$ and the $U(1)'$
$D$-term~\cite{Higgsbound}.
Higgs masses lighter than those allowed by LEP in the MSSM
are also possible, with the limits relaxed by mixings between
Higgs doublets and singlets.
There are stringent limits on an extra $Z'$ from direct searches
during Run I at the Tevatron~\cite{explim} and from indirect precision
 tests at the
$Z$-pole, at LEP 2, and from
 weak neutral current experiments~\cite{indirect}.
In typical models $M_{Z^{\prime}} > (500-800) $ GeV
and the $Z-\zpr$ mixing angle
$\alpha_{Z-Z^{\prime}}$ is smaller than a few $\x 10^{-3}$.
(The specific parameters considered here  yield a  $Z'$ mass of around
920 GeV, while the exclusion for  this case is $\sim$ 540 GeV.)
To explain the $Z-Z^{\prime}$ mass hierarchy without fine-tuning,
 two of us with J. Erler
 proposed a supersymmetric $U(1)^{\prime}$
model with a secluded $U(1)^{\prime}$-breaking
sector in which the squark and slepton spectra can mimic those of the MSSM,
the electroweak symmetry breaking is driven by relatively large $A$
terms, and a large \zpr \ mass can be generated by the VEVs of additional
SM singlet fields that are charged under the \upr \
but do not directly contribute to the effective $\mu$ parameter~\cite{ELL}.
Here, we consider EWBG in this model.

The model has one pair of
Higgs doublets $H_u$ and $H_d$, and four SM singlets, $S$, $S_1$,
$S_2$, and $S_3$ whose $U(1)'$ charges satisfy
\begin{eqnarray}
\label {qcharge} {Q_S=-Q_{S_1} =-Q_{S_2} ={1\over 2} Q_{S_3},~
Q_{H_d}+Q_{H_u}+Q_S=0.}
\end{eqnarray}
The superpotential for the Higgs sector is
\begin{eqnarray}
W_{H} &=& h S H_d H_u + \lambda S_1 S_2 S_3,\,
\end{eqnarray}
where $h$ is associated with the
effective $\mu$ term. The off-diagonal nature of $W_H$ is motivated
by string constructions.
We also introduce the supersymmetry breaking soft terms
\begin{eqnarray}
V_{soft}^{H} &=& m_{H_d}^2 |H_d|^2 + m_{H_u}^2 |H_u|^2 + m_S^2
|S|^2 + \sum_{i=1}^3 m_{S_i}^2 |S_i|^2 \nonumber\\&&
 -\left(A_h h S H_d
H_u + A_{\lambda} \lambda S_1 S_2 S_3
+ m_{S S_1}^2 S S_1
 \right.\nonumber\\&&\left.
 + m_{S S_2}^2 S S_2 + m_{S_1 S_2}^2 S_1^{\dagger} S_2
+ {\rm H. C.} \right)~.~\,
\label{vsoftH}
\end{eqnarray}
$m_{S S_1}^2$ and $m_{S S_2}^2$ are
 needed to break two unwanted global $U(1)$
symmetries. $m_{S_1 S_2}^2$ allows the possible tree-level
CP violation.
There is an almost $F$ and $D$ flat direction
involving $S_i$, with the flatness lifted by
 $\lambda$. For a sufficiently small $\lambda$,
the \zpr \ mass can be arbitrarily large\cite{ELL}.

An anomaly-free supersymmetric $U(1)'$ model can be constructed
by embedding
$SU(3)_C\times SU(2)_L \times U(1)_Y \times U(1)'$  into
 $E_6$. (We are not considering a full $E_6$ grand unified theory,
 but only using the \upr \ charges.)
 The $U(1)'$ is a linear combination of $U(1)_{\chi}$ and
$U(1)_{\psi}$,
 \begin{eqnarray}
Q^{\prime} &=& \cos\theta \ Q_{\chi} + \sin\theta \ Q_{\psi}~,~\,
\end{eqnarray}
where $U(1)_{\chi,\psi}$ are defined by
\begin{eqnarray}
E_6 \to\ SO(10) \x\ U(1)_{\psi} \to\ SU(5) \x\ U(1)_{\chi} \x\
U(1)_{\psi}~,~\,
\end{eqnarray}
and $\theta$ is a mixing angle.
We assume that the orthogonal $U(1)'$ is absent or very heavy.
We assume three ${\bf 27}$s, which include
 three families of the SM fermions, one
pair of Higgs doublets ($H_u$ and $H_d$), and a number of
SM singlets, extra Higgs-like doublets, and other exotics.
We assume that
the four SM singlets $S$, $S_1$, $S_2$, $S_3$
are the $S_L$, $S_L^*$, $S_L^*$ and ${\bar N}^*$, respectively, in
two (partial) pairs of $\bf 27$ and $\bf 27^*$
which include the extra ${S_L}$ and ${\bar N}$
to cancel the $U(1)'$ anomalies.
To be consistent with minimal gauge coupling unification, we
introduce one pair of vector-like doublets
$H_u^{\prime}$ and $\bar H_u^{\prime}$ from a pair of
 $\bf 27+27^*$~\cite{PLJW}.
We assume the other particles in the  $\bf 27+27^*$ pairs
are very heavy.
From $Q_S = {1 \over 2} {Q_S}_3$, {\it i.e.}, $Q_{S_L}= {1
\over 2} Q_{{\bar N}^*}$, we obtain $\tan \theta = {\sqrt{15}
\over 9}$. The $U(1)_{\chi}$, $U(1)_{\psi}$ and $U(1)'$ charges
are given in Table 1. The tiny neutrino masses
can be generated, {\it e.g.},  by the double see-saw or Type II 
see-saw mechanisms.

\vspace*{2mm}
\noindent
{\small
Table\,1. Decomposition of the $E_6$ fundamental  ${\bf 27}$ representation
for the left-chiral fields
under $SO(10)$, $SU(5)$, and the $U(1)_{\chi}$, $U(1)_{\psi}$ and
$U(1)'$ charges.
}
\begin{center}
\setlength{\tabcolsep}{0.2pc}
\begin{tabular}{|c| c| c| c| c|}
\hline
\hline
&&&&\\[-3.0mm]
 $SO(10)$ & $SU(5)$ & $2 \sqrt{10} Q_{\chi}$ & $2 \sqrt{6}
Q_{\psi}$ & $2 \sqrt{15} Q'$  \\
\hline
16   &   $10~ (u,d,{\bar u}, {\bar e} )$ & -1 & 1  & $-{1/2}$ \, \\
            &   ${\bar 5}~ ( \bar d, \nu ,e)$  & 3  & 1  & 4          \\
            &   $1 \bar N$             & -5 & 1  & -5         \\
\hline
       10   &   $5~(D,H^{\prime}_u)$    & 2  & -2 & 1          \\
            &   ${\bar 5} ~(\bar D, H^{\prime}_d)$ & -2 &-2 & $-{7/2}$ \\
\hline
      1    &   $1~ S_L$                  &  0 & 4 & $5/2$ \\
\hline\hline
\end{tabular}
\end{center}

To study the electroweak phase transition and
baryogenesis, we need  the one-loop effective
potential at finite temperature.
In the 't~Hooft-Landau gauge and in the
$\overline{MS}$-scheme,
 it is~\cite{Quiros}
\begin{eqnarray}
\label{total}
V_{\rm{e}}(\phi,T) &=& V_0(\phi) + V_1(\phi,0) +
\Delta V_1(\phi,T) +\Delta V_{\rm{d}}(\phi,T),
\end{eqnarray}
where $V_0(\phi)$ is the tree-level potential, $V_1(\phi,0)$ and
$\Delta V_1(\phi,T)$ are the one-loop corrections at zero and
finite temperatures, and $\Delta V_{\rm{d}}(\phi,T)$ is the
multi-loop daisy correction. 
The technical and numerical details are given in Ref.~\cite{KLLL}.

The \upr \ is broken at a first phase transition around 1 TeV, with the
$S_i$ acquiring large VEVs and $S$ a much smaller one, and the
electroweak symmetry at a second transition at the critical temperature $T_c$.
We plot the potential versus the vacuum expectation value (VEV) 
$v\equiv\sqrt{|\langle H_u^0 \rangle|^2+|\langle H_d^0 \rangle|^2}$
by connecting the true and false minima
 for the temperatures near the EWPT
in FIG. 1
for a set of typical input parameters given in Table 2.
The
EWPT occurs at $T_c=120$ GeV, with
\begin{eqnarray}
v(T_c)/T_c=1.31 ~,~\,
\end{eqnarray}
 strong enough for EWBG.
The transition is induced by the trilinear term
$A_h h S H_d H_u$,
so, unlike in the MSSM, there is no upper bound on the lightest stop mass.

\begin{figure}
\begin{center}
\includegraphics[scale=0.75]{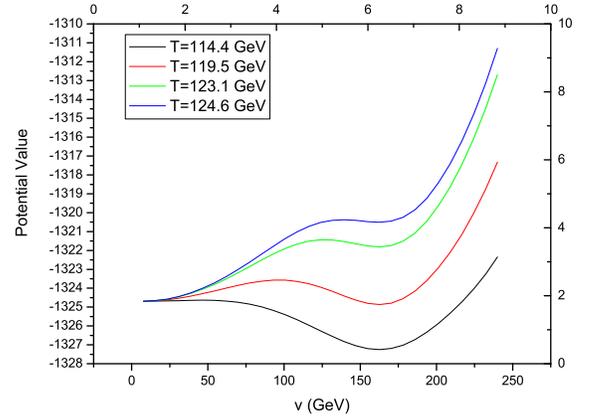}
\end{center}
\caption{The potential versus the VEV 
$v$ by connecting the true and false minima
where $v\equiv\sqrt{|\langle H_u^0 \rangle|^2+|\langle H_d^0 \rangle|^2}$.}
 \label{pothhs}
\end{figure}

\vspace*{2mm}
\noindent
{\small
Table\,2. A set of typical parameters. The energy unit  is 72 GeV.}
\begin{center}
\setlength{\tabcolsep}{0.2pc}
\begin{tabular}{|c|c|c|c|c|c|c|}
\hline
\hline
$h$ & $A_h$ & $\lambda$  & $A_{\lambda}$
& $m_{SS_1}^2$& $m_{SS_2}^2$ & $m_{S_1S_2}^2$  \\
\hline 0.8 & 3.2 & 0.06 & 3.1
& 0.012& 0.09 & 0.0003  \\
\hline $T$ & $m_{H_d^0}^2$ & $m_{H_u^0}^2$  & $m_S^2$
& $m_{S_1}^2$& $m_{S_2}^2$ & $m_{S_3}^2$ \\
\hline 1.65 & 0.25 & 0.25 & -0.25
& 0.031& 0.031 & -0.01 \\
\hline\hline
\end{tabular}
\end{center}

 The first order EWPT is realized by bubble nucleation. For
the non-local EWBG scenario the CP violation is associated with
particles interacting in the wall, and $B$ violation occurs near
the wall in the unbroken vacuum. Calculations have been
carried out in the thin and thick wall approximations~\cite{THICKWALL,CKNJPT}.
We will assume the thin wall regime which has a
relatively simple physical picture. This
is justified in our case for the $\tau$ lepton contribution
because the wall thickness $\delta$ is so small in comparison
with the $\tau$ mean free path. Using
the delta function type CP-violation source~\cite{KLLL,CKNJPT}, we have
\begin{eqnarray}
{n_B \over s}=-45(1 + {{v_w}\over {\langle v_L \rangle}}){\xi_L{m^2\delta
\Delta\theta_{CP}h(\delta,T)\Gamma_s^{'}}\over{(2\pi)^4v_w g_{*}T^4}},
\end{eqnarray}
where $m$, $D$, $\langle v_L \rangle$, and $\xi_L$ are respectively the fermion mass,
diffusion constant, average velocity, and persistence length in the wall frame;
$\Delta\theta_{CP}$ is the $CP$ phase change of the Higgs field coupling to the
fermion; $\Gamma_s^{'}$ is the sphaleron rate; and $h(\delta, T) \approx \int_0^T
dk_\bot \int_0^{1/\delta} dk_z{{k_\bot k_z}\over{\sqrt{k_\bot^2+k_z^2}}}$ is an
integral over momenta for the non-WKB fermions. Up to the leading order contribution for
the wall velocity $v_w$, Lorentz factors can be neglected in these formulae. Thus, three
parameters $\delta$, $v_w$ and $\Delta\theta_{CP}$ are the most important physical
quantities influencing EWBG.


The bubble wall can be approximated by the stationary
solution to the equations of motion for the Higgs fields.
 Solving these equations analytically is difficult due to
the Higgs field multiplicity. However, these field equations are
solved for the field configuration for which
\begin{eqnarray}
S_A = \int_{-\infty}^{+\infty}dz[\Sigma_iE^i_{m}(z)^2
+\Sigma_jE^j_{p}(z)^2] \equiv 0,
\end{eqnarray}
where $E^i_{m}(z)$ and $E^j_{p}(z)$ are the field equations for
the magnitude and phase, respectively ~\cite{MQSJCMS}. We
approximate  the wall profile and estimate $\delta$ by applying a
kink ansatz to minimize the action $S_A$. In our model, the
numerical results
show that the wall thickness $\delta$  varies
from $1 T^{-1}$ to $20 T^{-1}$ as a monotonically increasing
function of these phase changes, and is smaller
than the leptons' large mean free
path $\sim (30-70) T^{-1}$~\cite{CKNJPT}.


The wall velocity for a newly-born bubble is mainly
determined by the effective potential difference,
surface tension, and plasma friction. Analytic study
 is very difficult. Recent numerical
work~\cite{MJS} indicates that in the MSSM the wall is
extremely non-relativistic and can be as slow as $v_w \sim 0.01$.
For the $U(1)'$ model, a systematic study is absent.
However, there is a larger wall tension due to the TeV scale
$|S_i|$ and their space varying phases, and
 an associated larger critical radius (so the shrinking force decreases
more slowly). Meanwhile, the plasma friction is larger due to the
exotic particles. Thus, the wall velocity is slower than that in
the MSSM or NMSSM under the thin wall approximation, and it should
be extremely non-relativistic. This fact, together with
$\xi_L>30T^{-1}\gg \delta$ for left-chiral leptons in our model,
make the delta function type CP-violation source a good approximation
where no perturbation is generated behind the wall by the injected
current~\cite{CKNJPT}. To avoid unnecessary complication, we
still take $\xi_L=6D\langle v_L \rangle$ as in~\cite{CKNJPT}.


Only four combinations of the six
phases of the neutral Higgs fields are physical
\begin{eqnarray}
&&\beta_1=\theta_S+\theta_{S_1}~,~\beta_2=\theta_S+\theta_{S_2},\nonumber\\
&&\beta_3=\theta_S+\theta_{H_d^0} +\theta_{H_u^0}
~,~\beta_4=\theta_{S_1} +\theta_{S_2} +\theta_{S_3}  ~,~\,
\label{beta definition}
\end{eqnarray}
where $\theta_{\phi_i}$
is the phase of $\phi_i$.
The other two, $\Sigma_i Q_{\phi_i}^Z \theta_{\phi_i}$ and $
\Sigma_i Q_{\phi_i}' \theta_{\phi_i}$, where
$Q_{\phi_i}^Z$ and $Q_{\phi_i}'$ are respectively
the $U(1)_Z$ and
 $U(1)'$ charges, are gauge degrees of freedom.
 There are five complex parameters from the
 soft terms in Eq. (\ref{vsoftH}):
 $A_hh$, $A_\lambda\lambda$,
$m_{SS_1}^2$, $m_{SS_2}^2$ and $m_{S_1S_2}^2$.
Only four can be taken as real by field redefinition,
and thus, unlike the MSSM, complex VEVs of the Higgs and
singlet fields may be induced by an explicitly CP-violating phase $\gamma$ at
tree-level, where  $m^2_{S_1S_2}\equiv |m^2_{S_1S_2}|e^{i\gamma}$.
For the   relatively small values we choose for these soft masses,
the new contributions to the electric dipole
moments (EDMs) of the electron and neutron associated with this sector
(from Higgs scalar exchange) are five or six orders
smaller than the experimental bounds.

Unlike the MSSM,  spontaneous CP breaking (SCPB)
 can occur at tree-level  since
a mildly dominant $m_{S_1S_2}^2$ soft term will forbid the same
values for $\beta_1$ and $\beta_2$ for $\gamma= \pi$.
The SCPB
 only occurs in the domain
wall, and all gauge-independent phases are suppressed to zero in
the broken phase, so  new contributions to EDMs would vanish entirely.
However, bubbles of opposite
CP phase and baryon number may be produced, diluting the density.
(The explicit CP breaking from the fermion sector avoids problems
with cosmological domain walls~\cite{vilenkin}.) We therefore
choose $\gamma \ne \pi$.

The $CP$ phase change relevant to baryogenesis is
\begin{eqnarray}
\triangle \theta_{CP}=-{\frac{1}{5}}
\triangle \beta1 - {\frac{1}{5}} \triangle \beta_2 +
{\frac{7}{15}} \triangle \beta_3 + {\frac{2}{15}} \triangle
\beta_4.
\label{CP phase change}
\end{eqnarray}
The $| S_i|$ maintain almost the same large
values in both  phases,
leading to $\beta_4= \Delta\beta_4=0$.
However, $| S |$ changes, leading to changes in $\beta_{1,2,3}$.
We assume a kink ansatz for $\beta_{1,2}$ through the wall. However,
$\beta_3$ is tricky: from the soft term $A_h h S H_d^0 H_u^0$,
$\beta_3$ is suppressed to
zero once $H_d^0$ and $H_u^0$ obtain significant VEVs.
Due to loops, it is non-zero in the false vacuum, but numerical study shows
that the transition to zero occurs in the outer edge of the wall, where
$ H_{u,d}^0 $ are small.
In calculating the asymmetry, we consider the scattered
particles by the domain wall as freely propagating with
space-dependent mass $m_{\psi}(z) = h_{\psi} \frac{H_d^0(z)}{\sqrt
2}e^{i\Delta\theta_{H_d^0}(z)}$ for down-type fermions,
where $H_d^0(z)$ is also approximated by a kink function. $\beta_3$ is
therefore nonzero only when $H_d^0(z)$ is small and is irrelevant.
We therefore define an
effective CP phase change
\begin{eqnarray}
(\triangle \theta_{CP})_{eff} = -0.2(\triangle \beta_1
(z)+\triangle \beta_2(z))~.~\,
\end{eqnarray}

In FIG. 2, we show the $\gamma$ dependence of the baryon asymmetry
for the typical input parameters in Table 2 with $v_w=0.01$,
$0.02$ and $0.04$, respectively. We only consider the
contributions from the $\tau$ lepton because the quark diffusion
constants and light lepton masses are small. The $\tau$
lepton contribution is dominated by the left-chiral ones
 because the $\tau$ Yukawa coupling
is small due to $\tan\beta \sim 1$ in our model, which 
implies that the $\tau$ associated ``decay'' processes have no 
time to equilibrate in the right-chiral $\tau$
lepton diffusion tail~\cite{CKNJPT}, hence most of baryons 
are directly produced from the rejected left-chiral 
$\tau$ leptons. This plot shows
that, within theoretical uncertainties, the observed value can
be explained from the $\tau$ lepton contribution alone
 for $\gamma$ close to $\pi$. 
These results are rather conservative since we also neglect the
contributions from superparticles, such as charginos and
neutralinos.
 As a matter of fact, due to the space-dependent field
phases, even if $\tan\beta$ is a constant during EWPT and the
possible damping effects of particles in the bubble wall are also
counted in~\cite{Rius:1999zc}, 
the observed baryon asymmetry can be obtained
in our model through the chargino and neutralino contributions
if they are lighter than 800 GeV and the stop quarks are
not very heavy in comparison with $T_c$~\cite{KLLL}.

\begin{figure}
\begin{center}
\includegraphics[scale=0.45]{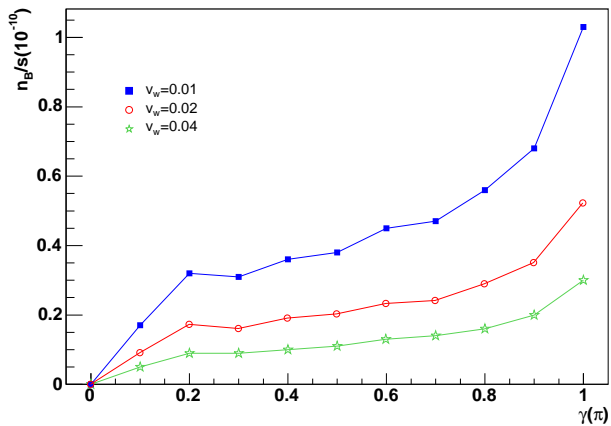}
\end{center}
\caption{Baryon asymmetry ($n_B/s$) vs. $\gamma$.}
\label{baryonasymmetrycurve}
\end{figure}


There is no baryon number dilution problem for $\gamma$ slightly
smaller than $\pi$, because the degenerate vacuum generic in SCPB
vanishes. There are degenerate false vacua differing in winding
number (i.e., phase changes of $2\pi$). However, these involve a
larger phase change with respect to the true vacuum, and their
bubble nucleation rate $\Gamma \sim
e^{-\int{dx^3\sum_{i=1}^3|S_i|^2({{d\theta_{S_i}}\over{dr}})^2/T}}$
is generally  negligible.

In summary, we considered the one-loop effective potential at finite
temperature in an anomaly free supersymmetric \upr \ model
with a secluded $U(1)^{\prime}$-breaking sector, and showed that there
exists a strong enough first order EWPT
for electroweak baryogenesis
due to the large trilinear term $A_h h S H_d H_u$.
 We briefly reviewed
the non-local electroweak baryogenesis mechanism (in the thin
wall regime), and found that within uncertainties
the observed baryon number can be generated
from the $\tau$ lepton contribution with the secluded
sector playing an essential role.

\smallskip

This research was supported in part by the U.S.~Department of
Energy under Grant No.~DOE-EY-76-02-3071 and by the National
Science Foundation under Grant No.~PHY-0070928.


\vspace*{-1.3mm}
\begin{thebibliography}{9}
\vspace*{-12mm}

\bibitem{WMAP}
D.~N.~Spergel {\it et al.},
Astrophys.\ J.\ Suppl.\  {\bf 148}, 175 (2003).

\bibitem{Sak}
A. D. Sakharov, JETP Lett. {\bf 5}, 24 (1967).

\bibitem{RTWB}
A.~Riotto and M.~Trodden,
Ann.\ Rev.\ Nucl.\ Part.\ Sci.\  {\bf 49}, 35 (1999);
W.~Bernreuther,
Lect.\ Notes Phys.\  {\bf 591}, 237 (2002), hep-ph/0205279;
and references therein.

\bibitem{MSSMEWBG}
B.~de Carlos and J.~R.~Espinosa, Nucl.\ Phys.\ B {\bf 503}, 24
(1997); M.~Carena, M.~Quiros and C.~E.~M.~Wagner,
 Phys.\ Lett. \ B {\bf 380}, 81 (1996);
M.~Carena, M.~Quiros, M.~Seco and C.~E.~M.~Wagner,
Nucl.\ Phys.\ B {\bf 650}, 24 (2003); K.~Kainulainen, T.~Prokopec,
M.~G.~Schmidt and S.~Weinstock, JHEP {\bf 0106}, 031 (2001).

\bibitem{Pietroni}
M.~Pietroni, Nucl.\ Phys.\ B {\bf 402}, 27 (1993).

\bibitem{DFMHS}
 A.~T.~Davies, C.~D.~Froggatt and R.~G.~Moorhouse,
Phys.\ Lett.\ B {\bf 372}, 88 (1996);
S. J.~Huber and M. G.~Schmidt,
 Eur. Phys. J. {\bf C10}, 473  (1999);
Nucl.\ Phys.\ B {\bf 606}, 183 (2001).

\bibitem{domain}
S.~A.~Abel, S.~Sarkar and P.~L.~White,
Nucl.\ Phys.\ B {\bf 454}, 663 (1995);
J.~Bagger, E.~Poppitz and L.~Randall,
Nucl.\ Phys.\ B {\bf 455}, 59 (1995), and references therein.

\bibitem{string}
M.~Cveti\v c and P.~Langacker,
Phys.\ Rev.\ D {\bf 54}, 3570 (1996) and
Mod.\ Phys.\ Lett.\ A {\bf 11}, 1247 (1996).

\bibitem{muprob1}
D.~Suematsu and Y.~Yamagishi,
Int.\ J.\ Mod.\ Phys.\ A {\bf 10}, 4521 (1995).

\bibitem{muprob2}
M.~Cveti\v c, D.~A.~Demir, J.~R.~Espinosa, L.~L.~Everett and P.~Langacker,
Phys.\ Rev.\ D {\bf 56}, 2861 (1997)
[Erratum-ibid.\ D {\bf 58}, 119905 (1997)].

\bibitem{Higgsbound}
L.~Durand and J.~L.~Lopez,
Phys.\ Lett.\ B {\bf 217}, 463 (1989);
M.~Drees, Int.\ J.\ Mod.\ Phys. A {\bf 4}, 3635 (1989);
M. Cveti\v c et al, ref.~\cite{muprob2}.

\bibitem{explim}  F.~Abe { et al.} [CDF Collaboration],
{ Phys.\ Rev.\ Lett.} {\bf 79}, 2192 (1997).

\bibitem{indirect}
J.~Erler and P.~Langacker,
Phys.\ Lett.\ B {\bf 456}, 68 (1999), and references therein.

\bibitem{ELL}
J.~Erler, P.~Langacker, T.~Li, Phys.\ Rev.\ D {\bf 66}, 015002
(2002).

\bibitem{PLJW}
P.~Langacker and J.~Wang,
Phys.\ Rev.\ D {\bf 58}, 115010 (1998).

\bibitem{Quiros}
M.~Quiros, hep-ph/9901312, and references therein.


\bibitem{KLLL}
J.~Kang, P.~Langacker, T.~Li and T.~Liu, in preparation.

\bibitem{THICKWALL}
M.~Carena, M.~Quiros, M.~Seco and C.~E.~M.~Wagner,
Nucl.\ Phys.\ B {\bf 650}, 24 (2003); K.~Kainulainen, T.~Prokopec,
M.~G.~Schmidt and S.~Weinstock, JHEP {\bf 0106}, 031 (2001);
hep-ph/0312110.

\bibitem{CKNJPT}
A.~Cohen, D.~Kaplan and A.~Nelson,  Nucl.\ Phys.\ B {\bf 349}, 727
(1991); M.~Joyce, T.~Prokopec and N.~Turok, phys.\ Lett.\ B {\bf
338}, 269 (1994); Phys.\ Rev.\ D {\bf 53}, 2930 (1996).

\bibitem{MQSJCMS}
J.~M.~Moreno, M.~Quiros and M.~Seco, Nucl.\ Phys.\ B {\bf 526},
489 (1998); P.~John, Phys.\ Lett.\ B {\bf 452}, 221 (1999);
J.~M.~Cline, G.~D.~Moore and G.~Servant, Phys.\ Rev.\ D {\bf 60},
105035 (1999).

\bibitem{MJS}
G.~D.~Moore, JHEP {\bf 0003}, 006 (2000); P.~John and
M.~G.~Schmidt, Nucl.\ Phys.\ B {\bf 598}, 291 (2001)
[Erratum-ibid.\ B {\bf 648}, 449 (2003)].

\bibitem{vilenkin}
A.~Vilenkin,
Phys.\ Rept.\  {\bf 121}, 263 (1985).

\bibitem{Rius:1999zc}
N.~Rius and V.~Sanz,
Nucl.\ Phys.\ B {\bf 570}, 155 (2000).


\end{thebibliography}
\end{document}